
\input epsf
\input phyzzx
\overfullrule=0pt

\def\half{{1\over 2}}
\def\p{\partial}

\def\scrim{{\cal I}^-}
\def\sp{\sigma^+}

\def\xp{x^+}
\def\xm{x^-}
\def\ap{a^+}
\def\am{a^-}
\def\omp{\p_+\Omega}

\def\ompm{\p_+\p_-\Omega}
\def\gl{\lambda}

\rightline{NSF-ITP-93-144}
\rightline{hep-th/9312017}

\vfill

\title{Universality and Scaling at the Onset of
Quantum Black Hole Formation}

\vfill

\author{Andrew Strominger,\foot{andy@denali.physics.ucsb.edu}}

\address{Department of Physics \break University of California at
\break Santa Barbara, CA 93106-9530}

\andauthor{L\'arus Thorlacius,\foot{larus@nsfitp.itp.ucsb.edu}}

\address{Institute for Theoretical Physics
\break University of California at
\break Santa Barbara, CA 93106-4030}

\vfill

\abstract
{\singlespace
In certain two-dimensional models, collapsing matter forms a
black hole if and only if the incoming energy flux exceeds the
Hawking radiation rate.  Near the critical threshold, the black
hole mass is given by a universal formula in terms of the
distance from criticality, and there exists a scaling solution
describing the formation and evaporation of an arbitrarily
small black hole.}

\vfill\endpage

\REF\choptuik{M.~W.~Choptuik
\journal Phys. Rev. Lett. & 70 (93) 9.}

\REF\chrs{D.~Christodolou
\journal Comm. Math. Phys. & 105 (86) 337;
{\bf 106} (1986), 587; {\bf 109} (1987) 613;
A.~M.~Abrahams and C.~R.~Evans
\journal Phys. Rev. Lett. & 70 (93) 2980.}

\REF\rev{For reviews see: J.~A.~Harvey and A.~Strominger,
in the Proceedings of the TASI Summer School, June 3-28, 1992,
Boulder, Colorado, (World Scientific, 1993);  S.~B.~Giddings,
in the Proceedings of the International Workshop on Theoretical
Physics, 6th Session, June 21-28, 1992, Erice, Italy,
(World Scientific, 1993).}

\REF\rst{J.~G.~Russo, L.~Susskind, and L.~Thorlacius
\journal Phys. Rev. & D46 (92) 3444
\journal Phys. Rev. & D47 (93) 533.}


\REF\asspt{A. Strominger and S. Trivedi, unpublished.}

\REF\cpt{A.~Strominger,\journal Phys. Rev. & D48 (93) 5769. }

\REF\cghs{C.~G.~Callan, S.~B.~Giddings, J.~A.~Harvey and
A.~Strominger
\journal Phys. Rev. & D45 (92) R1005.}

\REF\num{D. Lowe
\journal Phys. Rev. & D47 (93) 2446;
T. Piran and A.~Strominger
\journal Phys. Rev. & D48 (93) 4729.}

\REF\rstone{J.~G.~Russo, L.~Susskind, and L.~Thorlacius
\journal Phys. Lett. & B292 (92) 13.}

Recently remarkable critical behavior has been discovered at the
onset of classical black hole formation.  Choptuik [\choptuik]
considered a two-dimensional theory obtained from the $S$-wave
sector of four-dimensional general relativity coupled to a
massless scalar field. A sufficiently weak incoming $S$-wave pulse is
simply reflected through the origin to an outgoing pulse.
This behavior changes qualitatively as the initial amplitude
is increased.
Above a certain threshold, the pulse crosses its own Schwarzschild
radius before it reaches the origin, and a black hole is formed.
Choptuik [\choptuik] numerically investigated the mass $M_{BH}$ of the
resulting black hole as a function of the distance $\delta$
in the initial data space from the threshold. For small $\delta$
he finds:
$$
\log{M_{BH}}=\gamma\, \log{\delta} +{\cal O}( \delta^0),
\eqn\chop
$$
where the critical exponent $\gamma$ is numerically found to be near
$.37$. This result appears quite universal and is insensitive to the
precise definition of $\delta$ or scalar field couplings. In addition
the near-threshold scaling solution was
found to have fascinating self-similar
oscillations.  Related work can be found in [\chrs].
At present there appears to be little analytic or conceptual
understanding of these interesting phenomena.

Two-dimensional theories obtained
by  $S$-wave reduction from four dimensions have also been of recent
interest as simplified arenas for the study of quantum black hole
evaporation [\rev].  In this context null matter which
reduces to a free conformal field in
two dimensions is usually considered
because it is much simpler than a scalar
field (which reduces to a field
with complicated gravitational couplings).
At the classical level, every incoming $S$-wave pulse of such null
matter, no matter how weak, forms a black hole.
A threshold appears, however, when quantum effects are incorporated:
Energy must be thrown at the origin at a rate faster than a
newly-formed black hole wants to Hawking evaporate.

In this paper we study the onset of quantum
black hole formation using the
semiclassically soluble two-dimensional RST model [\rst].
We find analytically that
$$
\log{M_{BH}\over \gl}=\half \log{\delta}
-\half \log{\alpha}
+{\cal O}( \delta^0) \>.
\eqn\qchop
$$
$M_{BH}$ is defined here as the incoming energy of the null matter
swallowed by the black hole during its lifetime and $\gl$ is the
dimensionful parameter of the model.  The ${\cal O}(\delta^0)$ term
depends universally on $\alpha$, the second derivative of the energy
density at the point where the
critical threshold is exceeded\foot{It
would be interesting to determine if the
${\cal O}(\delta^0)$ term of \chop\
has a similar universal dependence.}.
The ${\cal O}( \delta^{1/2})$ corrections
depend non-universally on the shape of the incoming pulse. We also
find a scaling solution near criticality which corresponds
to the formation and
evaporation of an arbitrarily small black hole.
We have not determined whether the scaling \qchop\ is universal
with respect to small changes in the
coupling constants of the theory.
Presumably numerical work is required to answer this interesting
question.

There are obvious similarities between our results and those of
Ref. [\choptuik], but there are also apparent differences.  First of
all, our critical exponent is a rational number whereas Choptuik's
at least appears to be irrational.  Second, there is no analog of the
self-similar oscillations in our work.  This could be a special
feature arising from the linear nature of the RST equations, while
more general two-dimensional models (which are not analytically
soluble) might exhibit such oscillations.

We now present a derivation of the scaling relation \qchop .
The semi-classical effective action for the
RST model\foot{For more details on this model consult [\rst,\cpt].
We use the conventions of [\cpt].} is
$$\eqalign{
S = {1 \over 2\pi} \int & d^2x \sqrt{-g}
[(e^{-2\phi}-{N\over24}\phi)R+
4e^{-2\phi}\bigl((\nabla \phi)^2 + \lambda^2\bigr)
- \half \sum_{i=1}^N (\nabla f_i)^2] \cr
&- {N \over 96\pi} \int
d^2x \sqrt{-g(x)} \int d^2x' \sqrt{-g(x')}
\> R(x) G(x;x') R(x')  \>,  \cr}
\eqn\effact
$$
where $g_{\mu\nu}$ is the two-dimensional metric, $\phi$ is a scalar
field called the dilaton, $f_i$ are $N$ minimally coupled scalar
matter fields and $G$ is a Green function for the operator $\nabla^2$.
The effective action includes
the one-loop Liouville term due to the matter fields and if $N$ is
large this term provides the dominant quantum back-reaction on the
geometry.  This model differs from the original CGHS model [\cghs]
by a finite local counterterm, which restores a global symmetry of the
classical theory and enables writing down exact semi-classical
solutions in a rather simple form.  Numerical analyses of the original
model [\num] indicate, however, that the two models are similar,
\ie\ that the qualitative behavior of semi-classical solutions is not
sensitive to the existence of the global symmetry.

It is convenient to work in conformal gauge,
$g_{+-}=-\half e^{2\rho}$, $g_{++}=g_{--}=0$,
and use the global symmetry to further fix the coordinates
to ``Kruskal gauge'', where
$\rho=\phi+\half \log{N\over 12}$.
This eliminates the conformal factor from the discussion.
If we define a new dilaton field,
$$
\Omega = {12\over N}e^{-2\phi}
+\half \phi+{1\over 4}\log{N\over 48} \>,
\eqn\redef
$$
the semi-classical equations reduce to
$$\eqalign{
\p_+\p_- f_i =&\> 0 \>, \cr
\ompm =& -\gl^2 \>, \cr
-\p^2_\pm\Omega =&\, T^f_{\pm\pm} + t_\pm \>, \cr}
\eqn\eom
$$
where $T^f_{\pm\pm}={6\over N}\sum_{i=1}^N (\p_\pm f_i)^2$.
The function $t_+(\xp)$ takes the value $t_+=-{1\over 4 (\xp)^2}$
in Kruskal coordinates for any incoming matter energy flux which
vanishes sufficiently rapidly at asymptotic early and late times.
The field redefinition \redef\ is degenerate at
$\Omega={1\over 4}$ and $\Omega < {1\over 4}$ does not correspond
to a real value of $\phi$.  The curve
$\Omega = {1\over 4}$
is the analog of the origin of radial coordinates in higher
dimensional gravity\foot{This analogy can be made precise when the
model is interpreted as an effective theory for radial modes of
near extreme magnetic dilaton black holes in four dimensional
gravity [\rev].} and solutions should not be continued beyond it.
Instead, RST impose the following boundary conditions at this
curve, wherever it is timelike [\rst],
$$
\p_\pm\Omega \Big|_{\Omega={1\over 4}} = 0 \>.
\eqn\bcond
$$
Incoming energy flux is then reflected off the boundary and the
total outgoing flux (including the anomalous part $t_-$) can be
determined by using \bcond .
The boundary curve undergoes dynamical motion in response to the
incoming matter and this gives rise to some non-trivial behavior
in this model, in spite of the extreme simplicity of the field
equations \eom .

The boundary conditions \bcond\ ensure semi-classical energy
conservation and also that the physical curvature remains finite
at the boundary curve as long as it is timelike.
It should be noted, however, that these boundary conditions are not
the most general ones allowed, and our results may depend
qualitatively on this choice. Indeed \bcond\ is incompatible
with Dirichlet or Neumann boundary conditons on the $f_i$,
and may not be realizable as the semiclassical limit of any
fully quantum mechanical boundary conditions [\asspt].
Alternate possibilities are currently being explored.

The solution corresponding to incoming matter energy flux, which
tapers off at early and late times, but is otherwise quite general,
is given by
$$
\Omega (\xp,\xm) = -\xp \bigl(\gl^2\xm +P_+(\xp)\bigr)
+{1\over \gl}M\bigl(\xp\bigr)
- {1\over \gl}M\bigl(\xp_B (\xm)\bigr)
-{1\over 4} \log{\bigl({\xp\over \xp_B (\xm)}\bigr)} \>,
\eqn\sol
$$
where
$$
M(\xp) = \gl \int_0^{x^+} du\, u\, T_{++}^f(u) \>,
\qquad
P_+(\xp) = \int_0^{\xp} du\, T_{++}^f(u)  \>,
\eqn\moments
$$
and $\xp_B (\xm)$ is the $\xp$ value of the point on the boundary
curve from which the reflected signal propagates to $(\xp,\xm)$.
Wherever the boundary curve is timelike its shape is a simple
function of the incoming energy flux,
$$
\gl^2\xm_B = -P_+(\xp_B)-{1\over 4\xp_B} \>.
\eqn\shape
$$
If, however, the incoming energy flux becomes larger than the
outgoing Hawking flux of a two-dimensional black hole,
$T^f_{++}(\xp)>{1\over 4(\xp)^2}$,\foot{The black hole temperature
is independent of mass in this model and an energy flux of
$T^f_{++}(\xp)=1/4(\xp)^2$ in Kruskal coordinates corresponds to a
uniform incoming flux, $T^f_{++}(\sp)=\gl^2/4$, in the coordinate,
$\gl\sp = \log{\gl\xp}$, appropriate to asymptotic inertial
observers.} for some value of $\xp$ then the boundary curve becomes
spacelike and it is less trivial to determine its shape.

Spacelike segments of the boundary curve are curvature singularities.
They form inside regions of future trapped points which are bounded
by an apparent horizon, located where $\p_+\Omega=0$ [\rstone].
Applying $\p_+$ to \sol\ one finds that the apparent horizon curve
$(\xp_H,\xm_H)$ satisfies \shape\ for all values of $\xp_H$.
In other words, the apparent horizon coincides with the boundary
curve where the latter is timelike, but where the boundary becomes
spacelike the two curves separate and the apparent horizon cloaks
the singularity, as shown in figure~1.  Once the incoming energy
flux falls below the threshold value the black hole evaporates and
the apparent horizon approaches the singularity. The curves join
again at the endpoint of the evaporation, which is denoted by $E$ in
figure~1.  The null line-segment
$\xm =x^-_E$, $\xp_B(\xm_E)<\xp<\xp_E$,
is the global event horizon of the geometry.  We define the black
hole mass to be the total energy (as measured at $\scrim$) which
enters the black hole
$$
M_{BH}= M\bigl(\xp_E\bigr)-M\bigl(\xp_B(\xm_E)\bigr) \>.
\eqn\massdef
$$
Since both ends of the global horizon are on the boundary curve and
also on the apparent horizon it
immediately follows from \sol\ and \shape\ that
$$
M_{BH} = {\gl\over 4}\,
\log{\Bigl({\xp_E\over \xp_B (\xm_E)}\Bigr)}\>.
\eqn\mass
$$

Now consider black hole formation just above threshold.  For
concreteness assume that the incoming energy flux has a maximum at
$\gl\sp=\log{\gl\xp}=0$, where its value is $T^f_{++}=\gl^2({1\over
4}+\delta)$, and that the flux is below threshold everywhere except near
$\sp=0$ so that only a single small black hole is formed.  In the scaling
limit, $\delta\rightarrow 0$, a generic incoming flux of this type may be
parametrized as follows near $\sp=0$,
$$
T^f_{++}(\sp)=\gl^2({1\over 4}
+\delta)(1-\alpha\, \gl^2{\sp}^2 +\ldots \bigr) \>.
\eqn\flux
$$
It turns out that the higher terms in the Taylor expansion of $T^f_{++}$ do not
contribute in the scaling limit.

\vskip 30pt
\vbox{
{\centerline{\epsfsize=3.0in \epsfbox{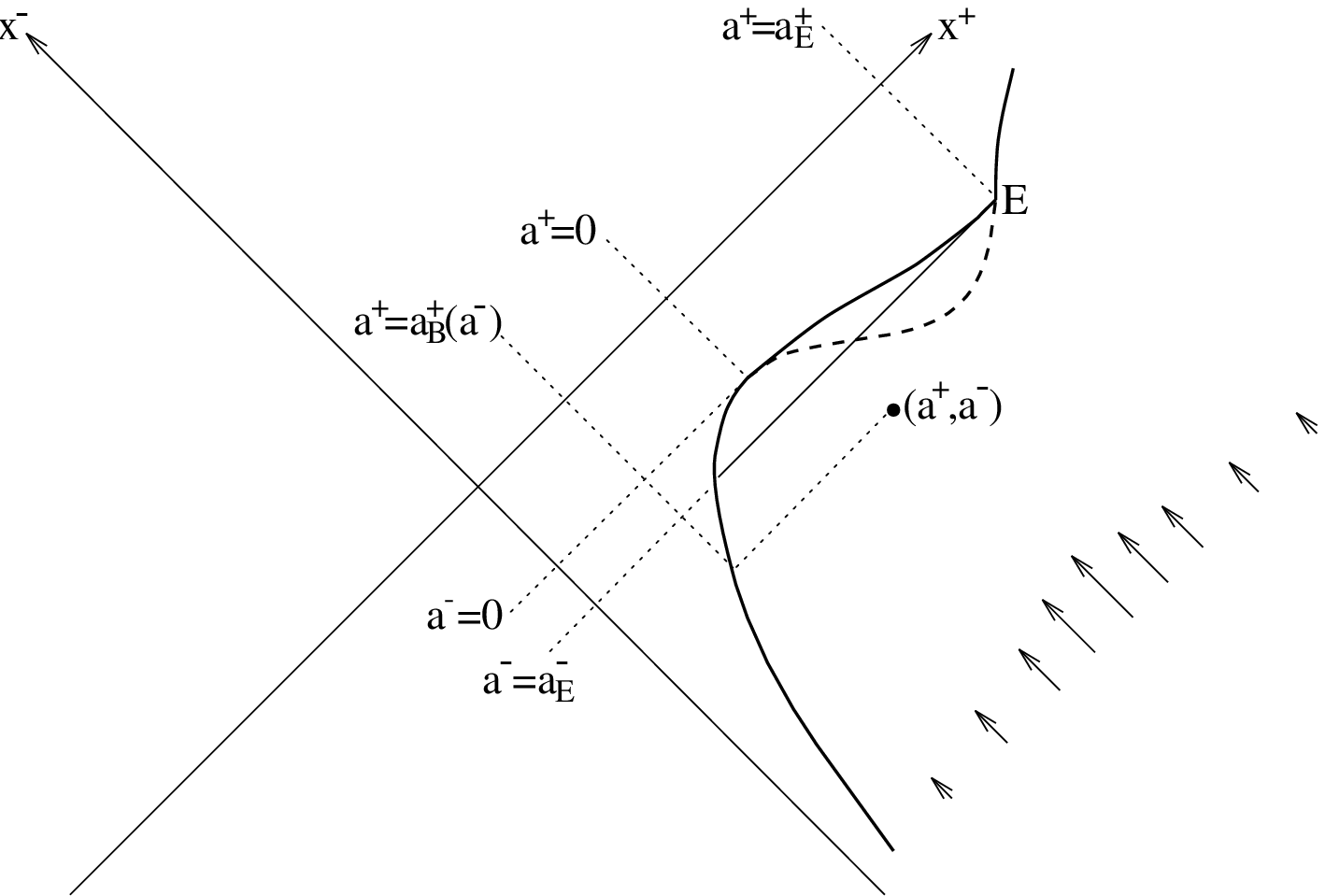}}}
\vskip 25pt
{\singlespace
{\centerline{\tenrm FIGURE 1. Kruskal diagram for black hole formation and
evaporation in the scaling}}
{\centerline{\tenrm regime.  A near-critical flux of matter energy is incident
from $\xm=-\infty$.  The solid}}
{\centerline{\tenrm curve is the $\Omega={1\over 4}$ boundary and the dashed
curve is the apparent horizon.}}
{\centerline{\tenrm The spacelike portion of the solid curve is the black hole
singularity.}}}
\vskip 15pt}

To obtain the shape of the apparent horizon curve, we transform to Kruskal
coordinates, compute $P_+(\xp)$ for this flux distribution, and insert the
result into \shape .  We are interested in the $\delta << 1$ limit and for that
purpose it is convenient to shift and rescale the Kruskal coordinates as
follows,
$$\gl\xp = 1+\sqrt{\delta\over \alpha}\,(\ap-2)
\>, \qquad\quad
\gl\xm = -{1\over \gl}P_+({1\over \gl})-{1\over 4}
+\sqrt{\delta^3\over \alpha}\,(\am+{4\over 3})   \>.
\eqn\abvar
$$
The origin of the $(\ap,\am)$ coordinate system has been chosen where, to
leading order in $\delta$, the boundary curve turns spacelike, as shown in
figure~1.  In the scaling region where higher order terms in $\delta$ can be
dropped, the apparent horizon takes a simple and universal form,
$$
\am_H(\ap_H) = -\half {\ap_H}^2 + {1\over 12} {\ap_H}^3\>.
\eqn\app
$$
The only reference made to the parameter $\alpha$ is in the definition of the
scaling variables \abvar .  Higher orders in the Taylor expansion of
$T^f_{++}(\sp)$ in \flux\ only contribute terms carrying positive powers of
$\delta$, which can be ignored in the scaling region.

It is straightforward to show that the singularity curve is also universal in
this limit.  Expressing \sol\ in terms of $(\ap,\am)$ coordinates and applying
${d\over d\ap}$ to both sides yields the following differential equation for
the $\Omega={1\over 4}$ curve,
$$
\bigl(\ap_S- \ap_B(\am_S)\bigr)\,
{d\am_S\over d\ap_S} =
-\am_S +\am_H(\ap_S) \>.
\eqn\singeq
$$
This is a non-linear differential equation and we do not have an analytic
solution, but there is no explicit dependence on the flux parameters so the
shape of the curve must be universal.    However, we do not need the whole
singularity curve in order to determine the black hole mass \mass .  It is
sufficient to locate the endpoint of evaporation and this can be achieved by
the following geometric argument, which is illustrated in figure~2.

\vskip 15pt
\vbox{
{\centerline{\epsfsize=3.0in \epsfbox{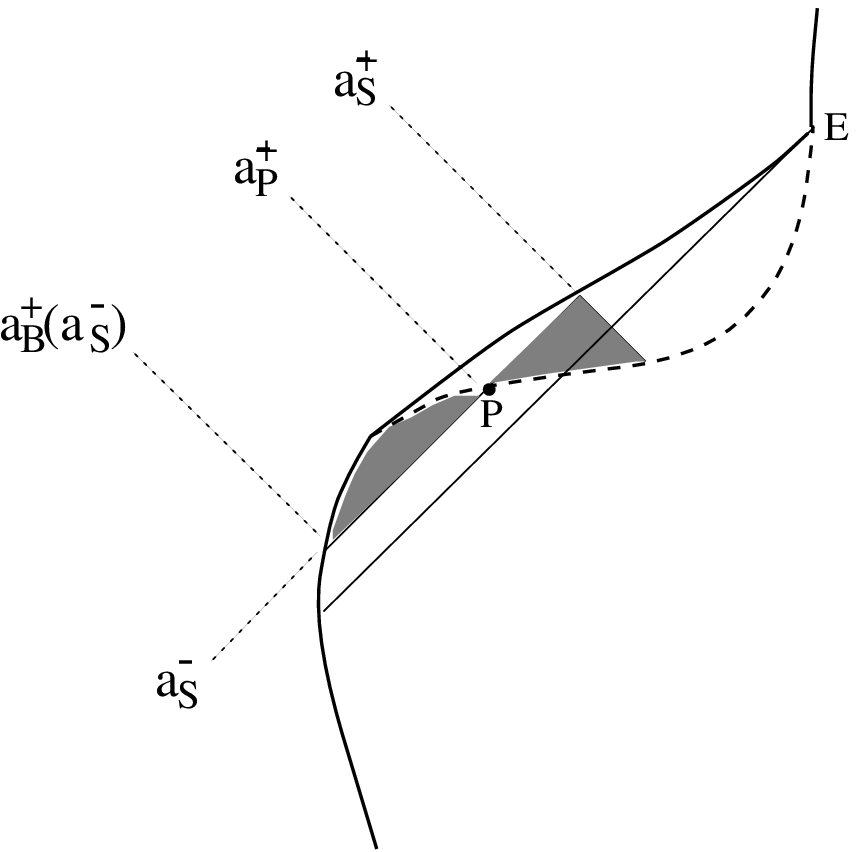}}}
\vskip 25pt
{\centerline{\tenrm FIGURE 2.  The two shaded regions have equal area
for any point $(\ap_S,\am_S)$ on the singularity curve.}}
\vskip 15pt}

Take any point on the singularity curve and consider the past-directed null
line $\am=\am_S$ which extends from $\ap=\ap_S$ to the timelike boundary curve
at $\ap=\ap_B(\am_S)$.  Denote by $P$ the point where this null line intersects
the apparent horizon.  The value of $\Omega(\ap_P,\am_P)$ can be obtained by
integrating $\omp$ along this null line starting either from the spacelike
singularity at $\ap_S$ or from the timelike boundary curve at $\ap_B(\am_S)$.
The value of $\omp$ on $\am=\am_S$ can in turn be obtained by integrating
$\p_-\p_+\Omega$ along null lines of constant $\ap$ starting from the apparent
horizon curve (where $\omp$ vanishes by definition).  By using the equation of
motion, $\p_-\p_+\Omega=-\gl^2$, inside the double integral one thus finds that
$\Omega(\ap_P,\am_P)-{1\over 4}$ equals the area of each of the shaded regions
in figure~2 and therefore these areas must be equal.  Interestingly, this
``equal area rule'' is an exact relationship which also holds outside of the
scaling limit.

The equal area rule, when applied at the global event horizon, $\am=\am_E$,
gives $\ap_E-2=2-\ap_B(\am_E)$.
Both endpoints of the global event horizon are also points on the apparent
horizon curve \app\ so we have
$-\half {\ap_E}^2+{1\over 12}{\ap_E}^3
=-\half \ap_B(\am_E)^2+{1\over 12}\ap_B(\am_E)^3$.
{}From these two relations it follows that $\ap_E=2+\sqrt{12}$,
$\ap_B(\am_E)=2-\sqrt{12}$, and the black hole mass \mass\ is to leading order
seen to be
$$
{M_{BH}\over \gl}=\sqrt{3\delta\over \alpha} \>,
\eqn\bhmass
$$
which is the scaling relation \qchop .

Finally, it is straightforward to integrate \singeq\ numerically to obtain the
entire singularity curve.  The result is plotted in figure~3.

\vskip 15pt
\vbox{
{\centerline{\epsfsize=3.0in \epsfbox{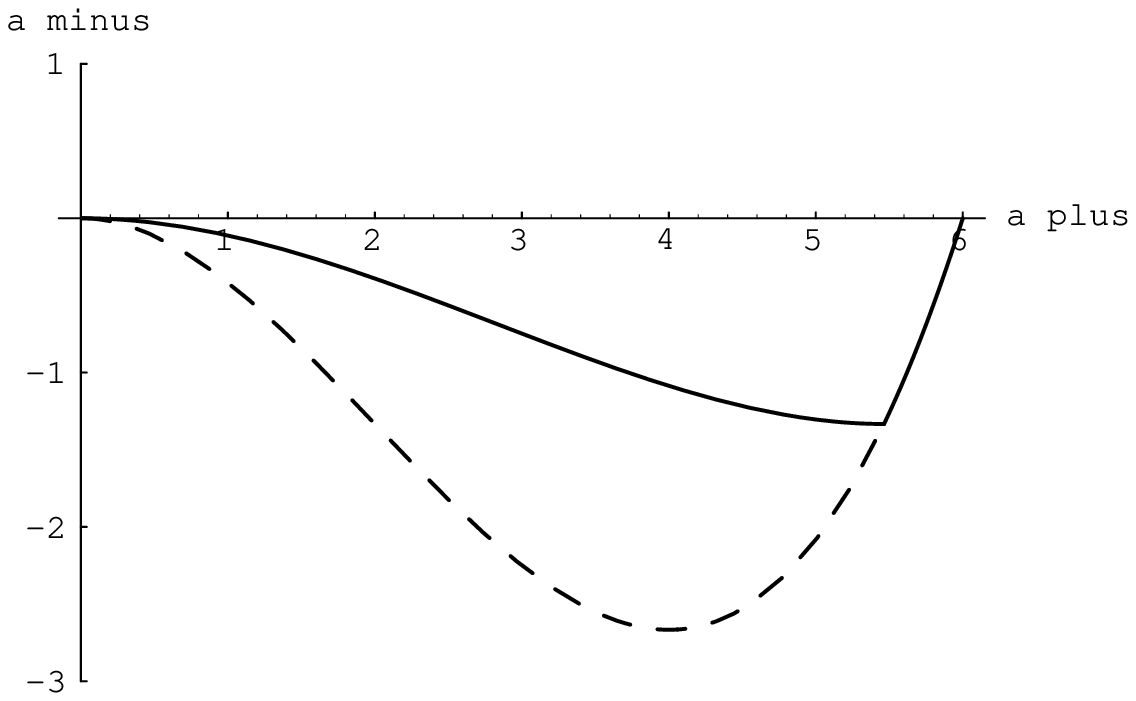}}}
{\singlespace
{\centerline{\tenrm FIGURE 3.  Singularity curve obtained from numerical
calculation.}}
{\centerline{\tenrm The dashed curve is the apparent horizon \app .}}}
\vskip 15pt}

\ack
This work was supported in part by DOE grant DOE-91ER40618 and NSF grant
PHY-89-04035.

\refout
\end